\documentclass[aps,prd,amssymb,amsmath,nofootinbib]{revtex4}

\begin{document}

\title{Fluctuations of an evaporating black hole from back reaction of
its Hawking radiation: Questioning a premise in earlier work}
\author{B.~L. Hu}
\author{Albert Roura}
\affiliation{Department of Physics, University of Maryland,
College Park, Maryland 20742-4111}

\begin{abstract}
This paper delineates the first steps in a systematic quantitative
study of the spacetime fluctuations induced by quantum fields in an
evaporating black hole. We explain how the stochastic gravity
formalism can be a useful tool for that purpose within a low-energy
effective field theory approach to quantum gravity. As an explicit
example we apply it to the study of the spherically-symmetric sector
of metric perturbations around an evaporating black hole background
geometry. For macroscopic black holes we find that those fluctuations
grow and eventually become important when considering sufficiently
long periods of time (of the order of the evaporation time), but well
before the Planckian regime is reached. In addition, the assumption of
a simple correlation between the fluctuations of the energy flux
crossing the horizon and far from it, which was made in earlier work
on spherically-symmetric induced fluctuations, is carefully analyzed
and found to be invalid. Our analysis suggests the existence of an
infinite amplitude for the fluctuations of the horizon as a
three-dimensional hypersurface. We emphasize the need for
understanding and designing operational ways of probing quantum metric
fluctuations near the horizon and extracting physically meaningful
information.
\end{abstract}


\maketitle

\centerline{\emph{Dedicated to Rafael Sorkin on the occasion of his 60th
birthday.}}

\section{Introduction}
\label{sec1}

Studying the dynamics of quantum fields in a fixed curved spacetime,
Hawking found that black holes emit thermal radiation with a
temperature inversely proportional to their mass \cite{hawking75}.
When the back reaction of the quantum fields on the spacetime dynamics
is included, one expects that the mass of the black hole decreases as
thermal radiation at higher and higher temperatures is emitted. This
picture, based on the process known as black hole evaporation, is
obtained from semiclassical gravity calculations which are believed to
be valid at least before the Planckian scale is reached
\cite{bardeen81,massar95}.

Semiclassical gravity \cite{birrell94,wald94,flanagan96} is a mean
field description that neglects the fluctuations of the spacetime
geometry. However, a number of studies have suggested the existence of
large fluctuations near black hole horizons
\cite{sorkin95,sorkin97,casher97,marolf03} (and even instabilities
\cite{mazur04}) with characteristic time-scales much shorter than the
black hole evaporation time. In all of them\footnote{At least those
which provide a relativistic description. The argument in
Refs.~\cite{sorkin95,sorkin97} is based on a non-relativistic
description and it is not obvious how to make some of our statements
precise in that case. However, a natural generalization to the
relativistic case is provided in Ref.~\cite{marolf03}, which does fall
into this category.} either states which are singular on the horizon
(such as the Boulware vacuum for Schwarzschild spacetime) were
explicitly considered, or fluctuations were computed with respect to
those states and found to be large near the horizon.  Whether these
huge fluctuations are of a generic nature or an artifact arising from
the consideration of states singular on the horizon is an issue that
deserves further investigation.  On the other hand, the fluctuations
for states regular on the horizon were estimated in Ref.~\cite{wu99}
and found to be small even when integrated over a time of the order of
the evaporation time. These apparently contradictory claims and the
fact that most claims on black hole horizon fluctuations were based on
qualitative arguments and/or semi-quantitative estimates prompted us
to strive for a more quantitative and self-consistent description.
(Previous attempts on this problem with similar emphasis by Raval,
Sinha and one of us have appeared in Refs.~\cite{hu99b,hu03b}. The
apparent difference between the conclusions in Ref.~\cite{hu99b} and
what is reported here will be explained below.)  In contrast to those
prior studies, we find here that in those cases the accumulated
fluctuations become significant by the time the black hole mass has
changed substantially, but well before reaching the Planckian regime.
Our result is in agreement with earlier work by
Bekenstein~\cite{bekenstein84}.

We will study metric fluctuations within a low-energy effective field
theory approach to quantum gravity \cite{burgess04}, which is commonly
believed to provide a valid description for phenomena involving
typical length-scales much larger than the Planck length even if the
microscopic details of the theory at Planckian scales are not known.
This approach has been mainly applied to weak field situations, such
as the study of quantum corrections to the Newtonian potential for
particles in a Minkowski background
\cite{donoghue94,donoghue97}. However, it is particularly interesting
to apply it also to strong field situations involving cosmological
\cite{weinberg05} or black hole spacetimes.  Here we will consider
quantized metric perturbations around a black hole background
geometry.

In the spirit of the effective field theory approach, the stochastic
gravity formalism \cite{hu03a,hu04a} provides a more than adequate and
perhaps the best available framework to study quantum metric
fluctuations because, for one reason at least, the correlation
functions that one obtains are equivalent to the quantum correlation
functions that would follow from a quantum field theory treatment, up
to a given order in an expansion in terms of the inverse number of
fields \cite{roura03b,hu04b}. Stochastic gravity allows for a
systematic study of the metric fluctuations in a black hole spacetime
over and beyond that for the mean value of the background spacetime
based on semiclassical gravity (with self-consistent back reaction
from the expectation value of the stress tensor). In tackling the
problem of metric fluctuations driven by quantum matter field
fluctuations for states regular on the horizon (as far as the
expectation value of the stress tensor is concerned), the existence of
correlations between the outgoing energy flux far from the horizon and
a negative energy flux crossing the horizon, based on energy
conservation arguments, constitutes an important step in previous
investigations \cite{bekenstein84,wu99} (see, however,
Refs.~\cite{parentani01b,parentani02}, where those correlators were
shown to vanish in an effectively two-dimensional model). Using
semiclassical gravity, such correlations have been confirmed for the
expectation value of the energy fluxes, provided that the mass of the
black hole is much larger than the Planck mass. However, a more
careful analysis, summarized in Sec.~\ref{sec4}, shows that no such
simple connection exists for energy flux fluctuations. It also reveals
that the fluctuations on the horizon, as a three-dimensional
hypersurface, are in fact divergent and one needs to find an adequate
way of probing the metric fluctuations near the horizon and extracting
physically meaningful information. The non-existence of this commonly
invoked relation in this whole subject matter illustrates the
limitations of heuristic arguments and the necessity of a detailed and
consistent formalism to study the fluctuations near the horizon.

We close this introduction with a few technical remarks. First, we
will restrict our attention to the spherically-symmetric sector of
metric fluctuations, which necessarily implies a partial description
of the fluctuations. Note, however, that contrary to the case for
semiclassical gravity solutions, even if one starts with
spherically-symmetric initial conditions, the stress tensor
fluctuations will induce fluctuations involving higher multipoles.
Thus, the multipole structure of the fluctuations is far richer than
that of spherically-symmetric semiclassical gravity solutions, but
this also means that obtaining a complete solution (including all
multipoles) for fluctuations rather than the mean value, even a
particular one, is much more difficult.

Second, for black hole masses much larger than the Planck mass
(otherwise the effective field theory description will break down
anyway), one can introduce a useful adiabatic approximation involving
inverse powers of the black hole mass. To obtain results to lowest
order, it is sufficient to compute the expectation value of the stress
tensor operator and its correlation functions in Schwarzschild
spacetime. The corrections proportional to higher powers of the
inverse mass can be neglected for sufficiently massive black holes.

Third, when studying the dynamics of induced metric fluctuations, the
additional contribution to the stress tensor expectation value which
results from evaluating it using the perturbed metric is often
neglected. In the consideration of fluctuations for an evaporating
black hole such a term (which will be denoted by $\langle
\hat{T}_{ab}^{(1)} [g+h] \rangle_\mathrm{ren}$ in Sec.~\ref{sec3})
becomes important when it builds up for long times. The importance of
this term is clear when comparing with the simple estimate made by Wu
and Ford in Ref.~\cite{wu99}, where $\langle \hat{T}_{ab}^{(1)} [g+h]
\rangle _\mathrm{ren}$ was neglected and the fluctuations were found
to be small even when integrated over long times, of the order of the
evaporation time of the black hole.

The paper is organized as follows. In Sec.~\ref{sec2} we briefly
review the results for the mean evolution of an evaporating black hole
obtained in the context of semiclassical gravity. The stochastic
gravity formalism is then applied in Sec.~\ref{sec3} to the study of
the spherically-symmetric sector of fluctuations around the
semiclassical gravity solution for an evaporating black hole.  The
existence of an exact correlation between the fluctuations of the
negative energy flux crossing the horizon and the flux far from it has
been previously assumed. In this paper we want to question this
assumption, but in the presentation in Sec.~\ref{sec3} we make such an
assumption to be in line with the literature. In Sec.~\ref{sec4}, we
present a careful analysis of this assumption, and show that it is
invalid. Further details can be found in a longer paper on the same
subject~\cite{hu06b}. Finally, in Sec.~\ref{sec5} we discuss several
implications of our results and suggest some directions for further
investigation.

Throughout the paper we use Planckian units with $\hbar=c=G=1$ and the
$(+,+,+)$ convention of Ref.~\cite{misner73}. We also make use of the
abstract index notation of Ref.~\cite{wald84}. Latin indices denote
abstract indices, whereas Greek indices are employed whenever a
particular coordinate system is considered.

\section{Mean evolution of an evaporating black hole}
\label{sec2}

\emph{Semiclassical gravity} provides a mean field description of
gravitational back-reaction problems in which the quantum effects of
the matter fields
can be important \cite{birrell94,wald94,flanagan96}. It is believed
to be applicable to situations involving length-scales much larger
than the Planck scale and for which the quantum back-reaction effects
due to the metric itself can be neglected as compared to those due to
the matter fields. In semiclassical gravity the spacetime geometry is
described by a classical metric while the matter fields are
quantized. The dynamics of the metric is governed by the semiclassical
Einstein equation:
\begin{equation}
G_{ab} \left[ g \right] = \kappa \left\langle
\hat{T}_{ab} [g] \right\rangle _\mathrm{ren}
\label{einstein1},
\end{equation}
where $\langle \hat{T}_{ab} [g] \rangle _\mathrm{ren}$ is the
renormalized expectation value of the stress tensor operator of the
quantum matter fields and $\kappa = 8 \pi / m_\mathrm{p}^2$ with
$m_\mathrm{p}^2$ being the Planck mass. Both the semiclassical
Einstein equation and the equation of motion for the matter fields
evolving in that geometry, whose solution is needed to evaluate
$\langle \hat{T}_{ab} [g] \rangle _\mathrm{ren}$, must be solved
self-consistently.

An important application of semiclassical gravity is the study of
black hole evaporation due to the back reaction of the Hawking
radiation emitted by the black hole on the spacetime geometry.
This has been studied in some detail for spherically symmetric black
holes \cite{bardeen81,massar95}. For a general spherically-symmetric
metric there always exists a system of coordinates in which it takes
the form
\begin{equation}
ds^2 = - e^{2 \psi(v,r)} ( 1 - 2 m(v,r)/r ) dv^2
+ 2 e^{\psi(v,r)} dv dr
+ r^2 \left( d\theta^2 + \sin^2 \theta d\varphi^2 \right)
\label{metric1}.
\end{equation}
This completely fixes the gauge freedom under local diffeomorphism
transformations except for an arbitrary function of $v$ that can be
added to the function $\psi(v,r)$ and is related to the freedom in
reparametrizing $v$ (we will see below how this can also be fixed).
In general this metric exhibits an \emph{apparent horizon}, where the
expansion of the outgoing radial null geodesics vanishes and which
separates regions with positive and negative expansion for those
geodesics, at those radii that correspond to (odd degree) zeroes of
the $vv$ metric component. Throughout the paper the location of the
apparent horizon will be denoted by $r_\mathrm{AH}(v)=2M(v)$, where
$M(v)$ satisfies the equation $2m(2M(v),v)=2M(v)$.

Spherical symmetry implies that the components $T_{\theta r}$,
$T_{\theta v}$, $T_{\varphi r}$ and $T_{\varphi v}$ vanish and the
remaining components are independent of the angular coordinates.
Under these conditions the various components of the semiclassical
Einstein equation associated with the metric in Eq.~(\ref{metric1})
become
\begin{eqnarray}
\frac{\partial m}{\partial v} &=& 4 \pi r^2 T_v^r
\label{einstein2a},\\
\frac{\partial m}{\partial r} &=& - 4 \pi r^2 T_v^v
\label{einstein2b},\\
\frac{\partial \psi}{\partial r} &=& 4 \pi r T_{rr}
\label{einstein2c},
\end{eqnarray}
where from now on we will simply use $T_{\mu \nu}$ to denote the
expectation value $\langle \hat{T}_{\mu \nu} [g] \rangle
_\mathrm{ren}$ and employ Planckian units (with $m_\mathrm{p}^2=1$).
Note that the arbitrariness in $\psi$ can be eliminated by choosing a
parametrization of $v$ such that $\psi$ takes a particular value at a
given radius (we will choose that it vanishes at $r=2M(v)$, where the
apparent horizon is located); $\psi$ is then entirely fixed by
Eq.~(\ref{einstein2c}).

Solving Eqs.~(\ref{einstein2a})-(\ref{einstein2c}) is no easy task.
However, one can introduce a useful adiabatic approximation in the
regime where the mass of the black hole is much larger than the Planck
mass, which is in any case a necessary condition for the semiclassical
treatment to be valid. What this entails is that when $M \gg 1$
(remember that we are using Planckian units) for each value of $v$ one
can simply substitute $T_{\mu \nu}$ by its ``parametric value'' -- by
this we mean the expectation value of the stress energy tensor of the
quantum field in a Schwarzschild black hole with a mass corresponding
to $M(v)$ evaluated at that value of $v$. This is in contrast to its
dynamical value, which should be determined by solving
self-consistently the semiclassical Einstein equation for the
spacetime metric and the equations of motion for the quantum matter
fields.
This kind of approximation introduces errors of higher order in
$L_\mathrm{H} \equiv B/M^2$ ($B$ is a dimensionless parameter that
depends on the number of massless fields and their spins and accounts
for their corresponding grey-body factors; it has been estimated to be
of order $10^{-4}$ \cite{page76}), which are very small for black
holes well above Planckian scales. These errors are due to the fact
that $M(v)$ is not constant and that, even for a constant $M(v)$, the
resulting static geometry is not exactly Schwarzschild because the
vacuum polarization of the quantum fields gives rise to a
non-vanishing $\langle \hat{T}_{ab} [g] \rangle _\mathrm{ren}$
\cite{york85}.

The expectation value of the stress tensor for Schwarzschild spacetime
has been found to correspond to a thermal flux of radiation (with
$T_v^r = L_\mathrm{H} / (4 \pi r^2)$) for large radii and of order
$L_\mathrm{H}$ near the horizon\footnote{The natural quantum state for
a black hole formed by gravitational collapse is the Unruh vacuum,
which corresponds to the absence of incoming radiation far from the
horizon. The expectation value of the stress tensor operator for that
state is finite on the future horizon of Schwarzschild, which is the
relevant one when identifying a region of the Schwarzschild geometry
with the spacetime outside the collapsing matter for a black hole
formed by gravitational collapse.}
\cite{candelas80,page82,howard84a,howard84b,anderson95}. This shows
the consistency of the adiabatic approximation for $L_\mathrm{H} \ll
1$: the right-hand side of Eqs.~(\ref{einstein2a})-(\ref{einstein2c})
contains terms of order $L_\mathrm{H}$ and higher, so that the
derivatives of $m(v,r)$ and $\psi(v,r)$ are indeed small.
Furthermore, one can use the $v$ component of the stress-energy
conservation equation (where terms of order $L_\mathrm{H}^2$ and
higher are neglected here)
\begin{equation}
\frac{\partial \left( r^2 T_v^r \right)}{\partial r}
+ r^2 \frac{\partial T_v^v}{\partial v} = 0
\label{conservation1},
\end{equation}
to relate the $T^r_v$ components on the horizon and far from it.
Integrating Eq.~(\ref{conservation1}) radially, one gets
\begin{equation}
(r^2 T^r_v) (r=2M(v),v) = (r^2 T^r_v) (r \approx 6M(v),v) +
O(L_{\mathrm{H}}^2), \label{conservation2}
\end{equation}
where we considered a radius sufficiently far from the horizon, but
not arbitrarily far (\emph{i.e.} $2M(v) \ll r \ll
M(v)/L_\mathrm{H}$). The second condition is necessary to ensure that
the size of the horizon has not changed much since the value of $v'$
at which the radiation crossing the sphere of radius $r$ at time $v$
left the region close to the horizon. Note that while in the nearly
flat region (for large radii) $T_v^r$ corresponds to minus the
outgoing energy flux crossing the sphere of radius $r$, on the
horizon, where $ds^2 = 2 e^{\psi(v,r)} dv dr + r^2 \left( d\theta^2 +
\sin^2 \theta d\varphi^2 \right)$, $T_v^r$ equals $T_{vv}$, which
corresponds to the null energy flux crossing the horizon. Hence,
Eq.~(\ref{conservation2}) relates the positive energy flux radiated
away far from the horizon and the negative energy flux crossing the
horizon. Taking into account this connection between energy fluxes and
evaluating Eq.~(\ref{einstein2a}) on the apparent horizon, we finally
get the equation governing the evolution of its size:
\begin{equation}
\frac{d M}{d v} = - \frac{B}{M^2}
\label{einstein3}.
\end{equation}
Unless $M(v)$ is constant, the event horizon and the apparent horizon
do not coincide. However, in the adiabatic regime their radii are
related, differing by a quantity of higher order in $L_\mathrm{H}$:
$r_\mathrm{EH}(v) = r_\mathrm{AH}(v) \, (1 + O(L_\mathrm{H}))$.

\section{Spherically-symmetric induced fluctuations}
\label{sec3}

There are situations in which the fluctuations of the stress tensor
operator and the metric fluctuations that they induce may be
important, so that the mean field description provided by
semiclassical gravity is incomplete and even fails to capture the most
relevant phenomena (the generation of primordial cosmological
perturbations constitutes a clear example of that). The
\emph{stochastic gravity} formalism \cite{hu03a,hu04a} provides a
framework to study those fluctuations. Its centerpiece is the
Einstein-Langevin equation
\begin{equation}
G_{ab}^{(1)}\left[ g+h\right] =\kappa \left\langle
\hat{T}_{ab}^{(1)} [g+h] \right\rangle _\mathrm{ren}
+\kappa \, \xi_{ab}\left[ g\right]
\label{einst-lang1},
\end{equation}
which governs the dynamics of the metric fluctuations around a
background metric $g_{ab}$ that corresponds to a given solution of
semiclassical gravity. The superindex $(1)$ indicates that only the
terms linear in the metric perturbations should be considered, and
$\xi_{ab}$ is a Gaussian stochastic source with vanishing expectation
value and correlation function\footnote{Throughout the paper we use
the notation $\langle \ldots \rangle_\xi$ for stochastic averages over
all possible realizations of the source $\xi_{ab}$ to distinguish them
from quantum averages, which are denoted by $\langle \ldots \rangle$.}
$\langle \xi_{ab} (x) \xi_{cd} (x') \rangle_\xi = (1/2) \langle \{
\hat{t}_{ab} (x), \hat{t}_{cd} (x') \} \rangle$ (with $\hat{t}_{ab}
\equiv \hat{T}_{ab} - \langle \hat{T}_{ab} \rangle$), where the term
on the right-hand side, which accounts for the stress tensor
fluctuations within this Gaussian approximation, is commonly known as
the noise kernel and denoted by $N_{abcd}(x,x')$. In this framework
the metric perturbations are still classical but
stochastic. Nevertheless, one can show that the correlation functions
for the metric perturbations that one obtains in stochastic gravity
are equivalent through order $1/N$ to the quantum correlation
functions that would follow from a quantum field theory treatment when
considering a large number of fields $N$ \cite{roura03b,hu04b}. In
particular, the symmetrized two-point function consists of two
contributions: \emph{intrinsic} and \emph{induced} fluctuations. The
intrinsic fluctuations are a consequence of the quantum width of the
initial state of the metric perturbations, and they are obtained in
stochastic gravity by averaging over the initial conditions for the
solutions of the homogeneous part of Eq.~(\ref{einst-lang1})
distributed according to
the reduced Wigner function associated with the initial quantum state
of the metric perturbations. On the other hand, the induced
fluctuations are due to the quantum fluctuations of the matter fields
interacting with the metric perturbations, and they are obtained by
solving the Einstein-Langevin equation using a retarded propagator
with vanishing initial conditions.

In this section, we will apply the stochastic gravity formalism to the
study of the spherically-symmetric sector (\emph{i.e.}, the monopole
contribution, which corresponds to $l=0$, in a multipole expansion in
terms of spherical harmonics $Y_{lm}(\theta,\phi)$) of metric
fluctuations for an evaporating black hole.
In this case only induced fluctuations are possible. The fact that
intrinsic fluctuations cannot exist can be clearly seen if one
neglects vacuum polarization effects, since Birkhoff's theorem
forbids the existence of spherically-symmetric free metric
perturbations outside the vacuum region of a spherically-symmetric
black hole that keep the ADM mass constant. Even when vacuum
polarization effects are included, spherically-symmetric
perturbations, characterized by $m(v,r)$ and $\psi(v,r)$, are not
independent degrees of freedom. This follows from
Eqs.~(\ref{einstein2a})-(\ref{einstein2c}), which can be regarded as
constraint equations.

The fluctuations of the stress tensor are inhomogeneous and
non-spherically-symmetric even if we choose a spherically-symmetric
vacuum state for the matter fields (spherical symmetry simply implies
that the angular dependence of the noise kernel in spherical
coordinates is entirely given by the relative angle between the
spacetime points $x$ and $y$). This means that, in contrast to the
semiclassical gravity case, projecting onto the $l = 0$ sector of
metric perturbations does not give an exact solution of the
Einstein-Langevin equation in the stochastic gravity approach that we
have adopted here. Nevertheless, restricting to spherical symmetry in
this way gives more accurate results than two-dimensional
dilaton-gravity models resulting from simple dimensional reduction
\cite{trivedi93,strominger93,lombardo99}. This is because we project
the solutions of the Einstein-Langevin equation just at the end,
rather than considering only the contribution of the $s$-wave modes to
the classical action for both the metric and the matter fields from
the very beginning. Hence, an infinite number of modes for the matter
fields with $l \neq 0$ contribute to the $l = 0$ projection of the
noise kernel, whereas only the $s$-wave modes for each matter field
would contribute to the noise kernel if dimensional reduction had been
imposed right from the start, as done in
Refs.~\cite{parentani01a,parentani01b,parentani02} as well as in
studies of two-dimensional dilaton-gravity models.

The Einstein-Langevin equation for the spherically-symmetric sector of
metric perturbations can be obtained by considering linear
perturbations of $m(v,r)$ and $\psi(v,r)$, projecting the stochastic
source that accounts for the stress tensor fluctuations to the $l=0$
sector, and adding it to the right-hand side of
Eqs.~(\ref{einstein2a})-(\ref{einstein2c}). We will focus our
attention on the equation for the evolution of $\eta(v,r)$, the
perturbation of $m(v,r)$:
\begin{equation}
\frac{\partial (m + \eta)}{\partial v} = - \frac{B}{(m + \eta)^2}
+ 4 \pi r^2 \xi_v^r + O \left(L_\mathrm{H}^2 \right)
\label{einst-lang2},
\end{equation}
which reduces, after neglecting terms of order $L_\mathrm{H}^2$ or
higher, to the following equation to linear order in $\eta$:
\begin{equation}
\frac{\partial \eta}{\partial v} = \frac{2 B}{m^3} \eta
+ 4 \pi r^2 \xi_v^r
\label{einst-lang3}.
\end{equation}
It is important to emphasize that in Eq.~(\ref{einst-lang2}) we
assumed that the change in time of $\eta(v,r)$ is sufficiently slow so
that the adiabatic approximation employed in the previous section to
obtain the mean evolution of $m(v,r)$ can also be applied to the
perturbed quantity $m(v,r)+\eta(v,r)$. This is guaranteed as long as
the term corresponding to the stochastic source is of order
$L_\mathrm{H}$ or higher, a point that will be discussed below.

Obtaining the noise kernel which determines the correlation function
for the stochastic source is highly nontrivial even if we compute it
on the Schwarzschild spacetime, which is justified in the adiabatic
regime for the background geometry. As implicitly done in prior work
(for instance in Refs.~\cite{bekenstein84,wu99}; see, however,
Refs.~\cite{parentani01b,parentani02}), we will assume in this section
that the fluctuations of the radiated energy flux far from the horizon
are exactly correlated with the fluctuations of the negative energy
flux crossing the horizon. This is a crucial assumption which implies
an enormous simplification and allows a direct comparison with the
results in the existing literature, and its validity will be analyzed
more carefully in the next section.\footnote{This simple relation
between the energy flux crossing the horizon and the flux far from it
is valid for the expectation value of the stress tensor, which is
based on an energy conservation argument for the $T_v^r$ component.
In most of the literature this relation is assumed to hold also for
fluctuations. However, in the next section we will show that this is
an incorrect assumption. Therefore, results derived from this
assumption and conclusions drawn are in principle suspect. (This
misstep is understandable because most authors have not acquired as
much insight into the nature of fluctuations phenomena as now.) Our
investigation testifies to the necessity of a complete reexamination
of all cases afresh. In fact, as we will show in the longer paper
following, an evaluation of the noise kernel near the horizon seems
unavoidable for the consideration of fluctuations and back-reaction
issues.}

Since the generation of Hawking radiation is especially sensitive to
what happens near the horizon, from now on we will concentrate on the
metric perturbations near the horizon\footnote{This means that
possible effects on the Hawking radiation due to the fluctuations of
the potential barrier for the radial mode functions will be missed by
our analysis.} and consider $\eta(v) = \eta(v,2M(v))$. Assuming that
the fluctuations of the energy flux crossing the horizon and those far
from it are exactly correlated, from Eq.~(\ref{einst-lang3}) we have
\begin{equation}
\frac{d \eta(v)}{d v} = \frac{2 B}{M^3(v)} \eta(v)
+ \xi(v)
\label{einst-lang4},
\end{equation}
where $\xi(v) \equiv (4 \pi r^2\, \xi_v^r) (v,r \approx 6M(v))$. The
correlation function for the spherically-symmetric fluctuation
$\xi(v)$ is determined by the integral over the whole solid angle of
the $N^{r\;r}_{\;v\;v}$ component of the noise kernel, which is given
by $(1/2) \langle \{ \hat{t}_v^r (x), \hat{t}_v^r (x') \}
\rangle$. The $l=0$ fluctuations of the energy flux of Hawking
radiation, characterized by $(1/2) \langle \{ \hat{t}_v^r (x),
\hat{t}_v^r (x') \} \rangle$, far from a black hole formed by
gravitational collapse have been studied in Ref.~\cite{wu99}. Its main
features are a correlation time of order $M$ and a characteristic
fluctuation amplitude of order $\epsilon_0 / M^4$ (this is the result
of smearing the stress tensor two-point function, which diverges in
the coincidence limit, over a period of time of the order of the
correlation time). The order of magnitude of $\epsilon_0$ has been
estimated to lie between $0.1 B$ and $B$ \cite{bekenstein84,wu99}. For
simplicity, we will consider quantities smeared over a time of order
$M$. We can then introduce the Markovian approximation $(\epsilon_0 /
M^3(v)) \delta(v-v')$, which coarse-grains the information on features
corresponding to time-scales shorter than the correlation time $M$.
Under those conditions $r^2 \xi^r_v$ is of order $1/M^2$ and the
adiabatic approximation made when deriving Eq.~(\ref{einst-lang2}) is
justified.

The stochastic equation (\ref{einst-lang4}) can be solved in the usual
way and the correlation function for $\eta(v)$ can then be computed.
Alternatively, one can follow Bekenstein \cite{bekenstein84} and
derive directly an equation for $\langle \eta^2 (v) \rangle_\xi$. This
is easily done multiplying Eq.~(\ref{einst-lang4}) by $\eta(v)$ and
taking the expectation value. The result is
\begin{equation}
\frac{d}{dv} \langle \eta^2 (v) \rangle_\xi
= \frac{4 B}{M^3(v)} \langle \eta^2 (v) \rangle_\xi
+ 2 \langle \eta (v) \xi (v) \rangle_\xi
\label{fluct1}.
\end{equation}
For delta-correlated noise (the Stratonovich prescription is the
appropriate one in this case), $\langle \eta (v) \xi (v) \rangle_\xi$
equals one half the time-dependent coefficient multiplying the delta
function $\delta (v-v')$ in the expression for $\langle \xi (v) \xi
(v') \rangle_\xi$, which is given by $\epsilon_0 / M^3(v)$ in our
case. Taking that into account, Eq.~(\ref{fluct1}) becomes
\begin{equation}
\frac{d}{dv} \langle \eta^2 (v) \rangle_\xi
= \frac{4 B}{M^3(v)}
\langle \eta^2 (v) \rangle_\xi + \frac{\epsilon_0}{M^3(v)}
\label{fluct2}.
\end{equation}
Finally, it is convenient to change from the $v$ coordinate to the
mass function $M(v)$ for the background solution. Eq.~(\ref{fluct2})
can then be rewritten as
\begin{equation}
\frac{d}{dM} \langle \eta^2 (M) \rangle_\xi
= - \frac{4}{M} \langle \eta^2 (M) \rangle_\xi
- \frac{(\epsilon_0 / B)}{M}
\label{fluct3}.
\end{equation}
The solutions of this equation are given by
\begin{equation}
\langle \eta^2 (M) \rangle_\xi
= \langle \eta^2 (M_0) \rangle_\xi \left(\frac{M_0}{M}\right)^4
+\frac{\epsilon_0}{4 B} \left[\left(\frac{M_0}{M}\right)^4 - 1\right]
\label{fluct4}.
\end{equation}
Provided that the fluctuations at the initial time corresponding to
$M=M_0$ are negligible (much smaller than $\sqrt{\epsilon_0 / 4B} \sim
1$), the fluctuations become comparable to the background solution
when $M \sim M_0^{2/3}$. Note that fluctuations of the horizon radius
of order one in Planckian units do not correspond to Planck scale
physics because near the horizon $\Delta R = r - 2M$ corresponds to a
physical distance $L \sim \sqrt{M \, \Delta R}$, as can be obtained
from the line element for Schwarzschild, $ds^2 = - (1-2M/r) dt^2 +
(1-2M/r)^{-1} dr^2 + r^2 (d\theta^2 + \sin^2 \theta d\varphi^2)$, by
considering pairs of points at constant $t$. So $\Delta R \sim 1$
corresponds to $L \sim \sqrt{M}$, whereas a physical distance of order
one is associated with $\Delta R \sim 1/M$, which corresponds to an
area change of order one for spheres with those radii. One can,
therefore, have initial fluctuations of the horizon radius of order
one for physical distances well above the Planck length provided that
we consider a black hole with a mass much larger than the Planck
mass. One expects that the fluctuations for states that are regular on
the horizon correspond to physical distances not much larger than the
Planck length, so that the horizon radius fluctuations would be much
smaller than one for sufficiently large black hole
masses. Nevertheless, that may not be the case when dealing with
states which are singular on the horizon, with estimated fluctuations
of order $M^{1/3}$ or even $\sqrt{M}$
\cite{casher97,marolf03,mazur04}. Confirming that the fluctuations are
indeed so small for regular states and verifying how generic, natural
and stable they are as compared to singular ones is a topic that we
plan to address in future investigations.

Our result for the growth of the fluctuations of the size of the black
hole horizon agrees with the result obtained by Bekenstein in
Ref.~\cite{bekenstein84} and implies that, for a sufficiently massive
black hole (with a few solar masses or a supermassive black hole), the
fluctuations become important before the Planckian regime is
reached. Strictly speaking, one cannot expect that a linear treatment
of the perturbations provides an accurate result when the fluctuations
become comparable to the mean value, but it signals a significant
growth of the fluctuations (at least until the nonlinear effects on
the perturbation dynamics become relevant).

This growth of the fluctuations which was found by Bekenstein and
confirmed here via the Einstein-Langevin equation seems to be in
conflict with the estimate given by Wu and Ford in
Ref.~\cite{wu99}. According to their estimate, the accumulated mass
fluctuations over a period of the order of the black hole evaporation
time ($\Delta t \sim M_0^3$) would be of the order of the Planck
mass. The discrepancy is due to the fact that the first term on the
right-hand side of Eq.~(\ref{einst-lang4}), which corresponds to the
perturbed expectation value $\langle \hat{T}_{ab}^{(1)} [g+h] \rangle
_\mathrm{ren}$ in Eq.~(\ref{einst-lang1}), was not taken into account
in Ref.~\cite{wu99}. The larger growth obtained here is a consequence
of the secular effect of that term, which builds up in time (slowly at
first, during most of the evaporation time, and becoming more
significant at late times when the mass has changed substantially) and
reflects the unstable nature of the background solution for an
evaporating black hole.\footnote{A clarification between our results
and the claims by Hu, Raval and Sinha in Ref.~\cite{hu99b} is in place
here: both use the stochastic gravity framework and perform an
analysis based on the Einstein-Langevin equation, so there should be
no discrepancy. However, the claim in Ref.~\cite{hu99b} was based on a
qualitative argument that focused on the stochastic source, but missed
the fact that the perturbations around the mean are unstable for an
evaporating black hole. Once this is taken into account, agreement
with the result obtained here is recovered.}

All this can be qualitatively understood as 
follows. Consider an evaporating black hole with initial mass $M_0$
and suppose that the initial mass is perturbed by an amount $\delta
M_0 = 1$. The mean evolution for the perturbed black hole (without
taking into account any fluctuations) leads to a mass perturbation
that grows like $\delta M = (M_0/M)^2 \, \delta M_0 = (M_0/M)^2$, so
that it becomes comparable to the unperturbed mass $M$ when $M \sim
M_0^{2/3}$, which coincides with the result obtained above. Such a
coincidence has a simple explanation: the fluctuations of the Hawking
flux slowly accumulated during most of the evaporating time, which are
of the order of the Planck mass, as found by Wu and Ford, give a
dispersion of that order for the mass distribution at the time when
unstable nature of the small perturbations around the background
solution start to become significant.

\section{Correlation between outgoing and ingoing energy fluxes}
\label{sec4}


In this section we will analyze more carefully whether the simple
relation between the energy flux crossing the horizon and the flux far
from it also holds for the fluctuations. One can find simple arguments
which show that those correlations vanish in two-dimensional
spacetimes \cite{hu06b}. Indeed, the correlation function for the
outgoing and ingoing null energy fluxes in an effectively
two-dimensional model was explicitly computed in
Refs.~\cite{parentani01b,parentani02} and found to vanish. On the
other hand, in four dimensions the correlation function does not
vanish in general and correlations between outgoing and ingoing fluxes
do exist near the horizon (at least partially). We plan to elaborate
further on these points in Ref.~\cite{hu06b}.

For black hole masses much larger than the Planck mass, one can use
the adiabatic approximation for the background mean
evolution. Therefore, to lowest order in $L_\mathrm{H}$ one can
compute the fluctuations of the stress tensor in Schwarzschild
spacetime. In Schwarzschild, the amplitude of the fluctuations of $r^2
\hat{T}^r_v$ far from the horizon is of order $1/M^2$ ($= M^2 / M^4$)
when smearing over a correlation time of order $M$, which one can
estimate for a hot thermal plasma in flat space
\cite{campos98,campos99}
(see Ref.~\cite{wu99} for a more accurate computation of the
fluctuations of $r^2 \hat{T}^r_v$ far from the horizon). The
amplitude of the fluctuations of $r^2 \hat{T}^r_v$ is thus of the same
order as its expectation value. However, their derivatives with
respect to $v$ are rather different: since the characteristic
variation times
for the expectation value and the fluctuations are $M^3$ and $M$
respectively, $\partial (r^2 T^r_v) / \partial v$ is of order $1/M^5$
whereas $\partial (r^2 \xi^r_v) / \partial v$ is of order $1/M^3$.
This implies an additional contribution of order $L_\mathrm{H}$ due to
the second term in Eq.~(\ref{conservation1}) if one radially
integrates the same equation applied to stress tensor fluctuations
(the stochastic source in the Einstein-Langevin equation). Hence, in
contrast to the case of the mean value, the contribution from the
second term in Eq.~(\ref{conservation1}) cannot be neglected when
radially integrating since it is of the same order as the
contributions from the first term, and one can no longer obtain a
simple relation between the outgoing energy flux far from the horizon
and the energy flux crossing the horizon.

So far we have argued that the method employed for the mean value
cannot be employed for the fluctuations. Although one expects that
$r^2 \xi^r_v$ on the horizon and far from it will not be equal when
including the contributions that results from radially integrating the
second term in Eq.~(\ref{conservation1}), one might wonder whether
there is a possibility that those contributions would somehow cancel
out. That possibility can, however, be excluded using the following
argument. The smeared correlation function
\begin{equation}
\int dv h(v) \int dv' h(v')\,
r^4 \langle \xi^r_v (v,r) \xi^r_v (v',r) \rangle_\xi
\label{correlation1},
\end{equation}
where $h(v)$ is some appropriate smearing function and $\xi^r_v (v,r)$
has already been integrated over the whole solid angle, is divergent
on the horizon but finite far from it. Therefore, $r^2 \xi^r_v$ on the
horizon and far from it cannot be equal for each value of $v$.

Let us discuss in some more detail the fact that certain smearings of
the quantity $r^4 \langle \xi^r_v (v,r) \xi^r_v (v',r) \rangle_\xi$
are divergent on the horizon but finite far from it.
The smeared correlation function is related to the noise kernel as follows:
\begin{equation}
\int dv dv' h(v) h(v')\,
r^4 \langle \xi^r_v (v,r) \xi^r_v (v',r) \rangle_\xi
= r^4 \int dv dv' h(v) h(v') \int d\Omega d\Omega'
N^{r\;r}_{\;v\;v} (v,r,\theta,\varphi;v',r,\theta',\varphi')
\label{correlation2}.
\end{equation}
The noise kernel is divergent in the coincident limit or for
null-separated points. Smearing the noise kernel along all directions
gives a finite result. However, although certain partial smearings
also give a finite result, others do not. For instance, smearing along
a timelike direction yields a finite result, whereas smearing on a
spacelike hypersurface yields in general a divergent result
\cite{ford05}. On the other hand, the result of smearing along two
``transverse'' null directions (two null directions sharing the same
orthogonal spacelike 2-surfaces) is also finite, but not for a
smearing along just one null direction even if we also smear along the
orthogonal spacelike directions. For $r>2M$ Eq.~(\ref{correlation2})
corresponds to a smearing along a timelike direction and gives a
finite result for the smeared correlation function, but on the horizon
it corresponds to a smearing along a single null direction and it is
divergent.

The proof of the results described in the previous paragraph will be
provided in Ref.~\cite{hu06b} by considering a smearing along all
directions and then taking the limit in which the smearing size along
one of the null directions vanishes. It proceeds in two steps. First,
it is shown in the flat space case. Then it is generalized to curved
spacetimes using a quasilocal expansion in terms of Riemann normal
coordinates.

\section{Discussion}
\label{sec5}

Using the stochastic gravity formalism, in Sec.~\ref{sec3} we found
that the spherically-symmetric fluctuations of the horizon size of
an evaporating black hole become important at late times, and even
comparable to its mean value when $M \sim M_0^{2/3}$, where $M_0$ is
the mass of the black hole at some initial time when the
fluctuations of the horizon radius are much smaller than the Planck
length.\footnote{Remember that for large black hole masses this can
still correspond to physical distances much larger than the Planck
length, as explained in Sec.~\ref{sec3}.} This is consistent with
the result previously obtained by Bekenstein in
Ref.~\cite{bekenstein84}.

It is important to realize that for a sufficiently massive black hole,
the fluctuations become significant well before the Planckian regime
is reached. More specifically, for a solar mass black hole they become
comparable to the mean value when the black hole radius is of the
order of $10 \mathrm{nm}$, whereas for a supermassive black hole with
$M \sim 10^7 M_\odot$, that happens when the radius reaches a size of
the order of $1 \mathrm{mm}$. One expects that in those circumstances
the low-energy effective field theory approach of stochastic gravity
should provide a reliable description.

Due to the nonlinear nature of the back-reaction equations, such as
Eq.~(\ref{einst-lang2}), the fact that the fluctuations can grow and
become comparable to the mean value implies non-negligible
corrections to the dynamics of the mean value itself. This can be
seen by expanding Eq.~(\ref{einst-lang2}) (evaluated on the horizon)
in powers of $\eta$ and taking the expectation value. Through order
$\eta^2$ we get
\begin{eqnarray}
\frac{d (M(v) + \langle \eta(v) \rangle_\xi)}{d v}
&=& - \left \langle \frac{B}{(M(v) + \eta(v))^2} \right \rangle_\xi
\nonumber \\
&=& - \frac{B}{M^2(v)} \left[ 1 - \frac{2}{M(v)} \langle \eta (v) \rangle_\xi
+ \frac{3}{M^2(v)} \langle \eta^2 (v) \rangle_\xi + O \left( \frac{\eta^3}{M^3}
\right) \right]  \label{rad_correction}.
\end{eqnarray}
When the fluctuations become comparable to the mass itself, the
third term (and higher order terms) on the right-hand side is no
longer negligible and we get non-trivial corrections to
Eq.~(\ref{einstein3}) for the dynamics of the mean value. These
corrections can be interpreted as higher order radiative corrections
to semiclassical gravity that include the effects of metric
fluctuations on the evolution of the mean value. For instance, the
third term on the right-hand side of Eq.~(\ref{rad_correction})
would correspond to a two-loop Feynman diagram involving a matter
loop with an internal propagator for the metric perturbations
(restricted to the spherically-symmetric sector in our case), as
compared to just one matter loop, which is all that semiclassical
gravity can account for.

An interesting aspect that we have not addressed in this work, but
which is worth investigating, is the quantum coherence of those
fluctuations. It seems likely that, given the long time periods
involved and the size of the fluctuations, the entanglement between
the Hawking radiation emitted and the black hole spacetime geometry
will effectively decohere the large horizon fluctuations, rendering
them equivalent to an incoherent statistical ensemble.


Does the existence of the significant deviations for the mean
evolution mentioned above imply that the results based on
semiclassical gravity obtained by Bardeen and Massar in
Refs.~\cite{bardeen81,massar95} are invalid? Several remarks are in
order. First of all, those deviations start to become significant only
after a period of the order of the evaporation time when the mass of
the black hole has decreased substantially. Secondly, since
fluctuations were not considered in those references, a direct
comparison cannot be established. However, we can compare the average
of the fluctuating ensemble with their results. Doing so exhibits an
evolution that deviates significantly when the fluctuations become
important. Nevertheless, if one considers a single member of the
ensemble at that time, its evolution will be accurately described by
the corresponding semiclassical gravity solution until the
fluctuations around that particular solution become important again,
after a period of the order of the evaporation time associated with
the new initial value of the mass at that time.


In this paper we take a first step to put the study of metric
fluctuations in black hole spacetimes on a firmer basis by considering
a detailed derivation of the results from an appropriate formalism
rather than using heuristic arguments or simple estimates.  The spirit
is somewhat analogous to the study of the mean back-reaction effect of
Hawking radiation on a black hole spacetime geometry (both for black
holes in equilibrium and for evaporating ones) by considering the
solutions of semiclassical gravity in that case rather than just
relying on simple energy conservation arguments. In order to obtain an
explicit result from the stochastic gravity approach and compare with
earlier work, in Sec.~\ref{sec3} we employed a simplifying assumption
implicitly made in most of the literature: the existence of a simple
connection between the outgoing energy flux fluctuations far from the
horizon and the negative energy flux fluctuations crossing the
horizon. In Sec.~\ref{sec4} we analyzed this assumption carefully and
showed it to be invalid. This strongly suggests that one needs to
study the stress tensor fluctuations from an explicit calculation of
the noise kernel near the horizon. This quantity is obtainable from
the stochastic gravity program and calculation is underway
\cite{phillips01,phillips03}.

A possible way to compute the noise kernel near the horizon could be
to use an approximation scheme based on a quasilocal expansion such as
Page's approximation \cite{page82} or similar methods corresponding to
higher order WKB expansions \cite{anderson95}.\footnote{Note, however,
that in most of these approaches the state of the quantum fields is
the Hartle-Hawking vacuum. For an evaporating black hole, one should
consider the Unruh vacuum.} With these techniques one can obtain an
approximate expression for the Wightmann function of the matter
fields, which is the essential object needed to compute the noise
kernel. Unfortunately these approximations are only accurate for pairs
of points with a small separation scale and break down when it becomes
comparable to the black hole radius.  Therefore, it cannot be employed
to study the $l=0$ multipole since that corresponds to averaging the
noise kernel over the whole solid angle, which involves typical
separations for pairs of points on the horizon of the order of the
black hole radius.
Alternatively, one might hope to gain some insight on the
fluctuations near a black hole horizon by studying the fluctuations
of the event horizon surrounding any geodesic observer in de Sitter
spacetime, which exhibits a number of similarities with the event
horizon of a black hole in equilibrium \cite{gibbons77a}. In
contrast to the black hole case, it may be possible to obtain exact
analytical results for de Sitter space due to its high degree of
symmetry.

Furthermore, as explained in Sec.~\ref{sec4} and shown in detail in
Ref.~\cite{hu06b}, the noise kernel smeared over the horizon is
divergent, and so are the induced metric fluctuations. Hence, one
cannot study the fluctuations of the horizon as a three-dimensional
hypersurface for each realization of the stochastic source because
they are infinite, even when restricting one's attention to the $l=0$
sector.  This means that one must find an adequate way of probing the
metric fluctuations and extracting physically meaningful information,
such as their effect on the Hawking radiation emitted by the black
hole. One possibility is to study how metric fluctuations affect the
propagation of a bundle of null geodesics
\cite{barrabes99,barrabes00,parentani01a,parentani01b,parentani02}. One
expects that this should provide a description of the effects on the
propagation of a test field whenever the geometrical optics
approximation is valid. However, if one tries to justify this point
starting with a quantum field theory treatment, one realizes that even
in simple cases interference effects cannot be neglected for
sufficiently long times (much longer than the inverse of the frequency
of the wave-packet whose propagation one is considering) and the
geometrical optics approximation is invalid. Another possibility,
which seems to constitute a better probe of the metric fluctuations,
is to analyze the effect on the two-point quantum correlation
functions of a test field. The two-point functions characterize the
response of a particle detector for that field and can be used to
obtain the expectation value and the fluctuations of the stress tensor
of the test field.

Finally, since the large fluctuations suggested in
Refs.~\cite{sorkin95,sorkin97,casher97,marolf03} involve time-scales
much shorter than the evaporation time (contrary to those considered
in this paper) and high multipoles, one expects that for a
sufficiently massive black hole the spacetime near the horizon can be
approximated by Rindler spacetime (identifying the black hole horizon
and the Rindler horizon) provided that we restrict ourselves to
sufficiently small angular scales. Thus, analyzing the effect of
including the interaction with the metric fluctuations on the
two-point functions of a test field propagating in flat space, which
is technically much simpler, could provide useful information for the
black hole case.

\begin{acknowledgments}
We thank Paul Anderson, Larry Ford, Valeri Frolov, Ted Jacobson,
Don Marolf, Emil Mottola, Don Page, Renaud Parentani and Rafael
Sorkin for useful discussions. This work is supported by NSF under
Grant PHY03-00710.
\end{acknowledgments}




\begin{thebibliography}{47}
\expandafter\ifx\csname natexlab\endcsname\relax\def\natexlab#1{#1}\fi
\expandafter\ifx\csname bibnamefont\endcsname\relax
  \def\bibnamefont#1{#1}\fi
\expandafter\ifx\csname bibfnamefont\endcsname\relax
  \def\bibfnamefont#1{#1}\fi
\expandafter\ifx\csname citenamefont\endcsname\relax
  \def\citenamefont#1{#1}\fi
\expandafter\ifx\csname url\endcsname\relax
  \def\url#1{\texttt{#1}}\fi
\expandafter\ifx\csname urlprefix\endcsname\relax\def\urlprefix{URL }\fi
\providecommand{\bibinfo}[2]{#2}
\providecommand{\eprint}[2][]{\url{#2}}

\bibitem[{\citenamefont{Hawking}(1975)}]{hawking75}
\bibinfo{author}{\bibfnamefont{S.~W.} \bibnamefont{Hawking}},
  \bibinfo{journal}{Comm. Math. Phys.} \textbf{\bibinfo{volume}{43}},
  \bibinfo{pages}{199} (\bibinfo{year}{1975}).

\bibitem[{\citenamefont{Bardeen}(1981)}]{bardeen81}
\bibinfo{author}{\bibfnamefont{J.~M.} \bibnamefont{Bardeen}},
  \bibinfo{journal}{Phys. Rev. Lett.} \textbf{\bibinfo{volume}{46}},
  \bibinfo{pages}{382} (\bibinfo{year}{1981}).

\bibitem[{\citenamefont{Massar}(1995)}]{massar95}
\bibinfo{author}{\bibfnamefont{S.}~\bibnamefont{Massar}},
  \bibinfo{journal}{Phys. Rev. D} \textbf{\bibinfo{volume}{52}},
  \bibinfo{pages}{5857} (\bibinfo{year}{1995}).

\bibitem[{\citenamefont{Birrell and Davies}(1994)}]{birrell94}
\bibinfo{author}{\bibfnamefont{N.~D.} \bibnamefont{Birrell}} \bibnamefont{and}
  \bibinfo{author}{\bibfnamefont{P.~C.~W.} \bibnamefont{Davies}},
  \emph{\bibinfo{title}{Quantum fields in curved space}}
  (\bibinfo{publisher}{Cambridge University Press},
  \bibinfo{address}{Cambridge}, \bibinfo{year}{1994}).

\bibitem[{\citenamefont{Wald}(1994)}]{wald94}
\bibinfo{author}{\bibfnamefont{R.~M.} \bibnamefont{Wald}},
  \emph{\bibinfo{title}{Quantum field theory in curved spacetime and black hole
  thermodynamics}} (\bibinfo{publisher}{The University of Chicago Press},
  \bibinfo{address}{Chicago}, \bibinfo{year}{1994}).

\bibitem[{\citenamefont{Flanagan and Wald}(1996)}]{flanagan96}
\bibinfo{author}{\bibfnamefont{E.~E.} \bibnamefont{Flanagan}} \bibnamefont{and}
  \bibinfo{author}{\bibfnamefont{R.~M.} \bibnamefont{Wald}},
  \bibinfo{journal}{Phys. Rev. D} \textbf{\bibinfo{volume}{54}},
  \bibinfo{pages}{6233} (\bibinfo{year}{1996}).

\bibitem[{\citenamefont{Sorkin}(1995)}]{sorkin95}
\bibinfo{author}{\bibfnamefont{R.~D.} \bibnamefont{Sorkin}}, in
  \emph{\bibinfo{booktitle}{Proceedings of the Conference on Heat Kernel
  Techniques and Quantum Gravity}}, edited by
  \bibinfo{editor}{\bibfnamefont{S.~A.} \bibnamefont{Fulling}}
  (\bibinfo{publisher}{University of Texas Press}, \bibinfo{address}{College
  Station, Texas}, \bibinfo{year}{1995}), Discourses in Mathematics and its
  Applications, vol. 4, \eprint{gr-qc/9508002}.

\bibitem[{\citenamefont{Sorkin}(1997)}]{sorkin97}
\bibinfo{author}{\bibfnamefont{R.~D.} \bibnamefont{Sorkin}}, in
  \emph{\bibinfo{booktitle}{Proceedings of the First Australasian Conference on
  General Relativity and Gravitation}}, edited by
  \bibinfo{editor}{\bibfnamefont{D.}~\bibnamefont{Wiltshire}}
  (\bibinfo{publisher}{University of Adelaide}, \bibinfo{address}{Adelaide,
  Australia}, \bibinfo{year}{1997}), \eprint{gr-qc/9701056}.

\bibitem[{\citenamefont{Casher et~al.}(1997)\citenamefont{Casher, Englert,
  Itzhaki, Massar, and Parentani}}]{casher97}
\bibinfo{author}{\bibfnamefont{A.}~\bibnamefont{Casher}},
  \bibinfo{author}{\bibfnamefont{F.}~\bibnamefont{Englert}},
  \bibinfo{author}{\bibfnamefont{N.}~\bibnamefont{Itzhaki}},
  \bibinfo{author}{\bibfnamefont{S.}~\bibnamefont{Massar}}, \bibnamefont{and}
  \bibinfo{author}{\bibfnamefont{R.}~\bibnamefont{Parentani}},
  \bibinfo{journal}{Nucl. Phys. B} \textbf{\bibinfo{volume}{484}},
  \bibinfo{pages}{419} (\bibinfo{year}{1997}).

\bibitem[{\citenamefont{Marolf}(2005)}]{marolf03}
\bibinfo{author}{\bibfnamefont{D.}~\bibnamefont{Marolf}}, in
  \emph{\bibinfo{booktitle}{Particle physics and the universe}}, edited by
  \bibinfo{editor}{\bibfnamefont{J.}~\bibnamefont{Trampetic}} \bibnamefont{and}
  \bibinfo{editor}{\bibfnamefont{J.}~\bibnamefont{Wess}}
  (\bibinfo{publisher}{Springer-Verlag}, \bibinfo{year}{2005}), Springer
  Proceedings in Physics, vol. 98, \eprint{hep-th/0312059}.

\bibitem[{\citenamefont{Mazur and Mottola}(2004)}]{mazur04}
\bibinfo{author}{\bibfnamefont{P.~O.} \bibnamefont{Mazur}} \bibnamefont{and}
  \bibinfo{author}{\bibfnamefont{E.}~\bibnamefont{Mottola}},
  \bibinfo{journal}{Proc. Nat. Acad. Sci.} \textbf{\bibinfo{volume}{111}},
  \bibinfo{pages}{9545} (\bibinfo{year}{2004}).

\bibitem[{\citenamefont{Wu and Ford}(1999)}]{wu99}
\bibinfo{author}{\bibfnamefont{C.~H.} \bibnamefont{Wu}} \bibnamefont{and}
  \bibinfo{author}{\bibfnamefont{L.~H.} \bibnamefont{Ford}},
  \bibinfo{journal}{Phys. Rev. D} \textbf{\bibinfo{volume}{60}},
  \bibinfo{pages}{104013} (\bibinfo{year}{1999}).

\bibitem[{\citenamefont{Hu et~al.}(1998)\citenamefont{Hu, Raval, and
  Sinha}}]{hu99b}
\bibinfo{author}{\bibfnamefont{B.~L.} \bibnamefont{Hu}},
  \bibinfo{author}{\bibfnamefont{A.}~\bibnamefont{Raval}}, \bibnamefont{and}
  \bibinfo{author}{\bibfnamefont{S.}~\bibnamefont{Sinha}}, in
  \emph{\bibinfo{booktitle}{Black Holes, Gravitational Radiation and the
  Universe: Essays in honor of C.~V.~Vishveshwara}}, edited by
  \bibinfo{editor}{\bibfnamefont{B.}~\bibnamefont{Iyer}} \bibnamefont{and}
  \bibinfo{editor}{\bibfnamefont{B.}~\bibnamefont{Bhawal}}
  (\bibinfo{publisher}{Kluwer Academic Publishers},
  \bibinfo{address}{Dordrecht}, \bibinfo{year}{1998}), \eprint{gr-qc/9901010}.

\bibitem[{\citenamefont{Sinha et~al.}(2003)\citenamefont{Sinha, Raval, and
  Hu}}]{hu03b}
\bibinfo{author}{\bibfnamefont{S.}~\bibnamefont{Sinha}},
  \bibinfo{author}{\bibfnamefont{A.}~\bibnamefont{Raval}}, \bibnamefont{and}
  \bibinfo{author}{\bibfnamefont{B.~L.} \bibnamefont{Hu}},
  \bibinfo{journal}{Found. Phys.} \textbf{\bibinfo{volume}{33}},
  \bibinfo{pages}{37} (\bibinfo{year}{2003}).

\bibitem[{\citenamefont{Bekenstein}(1984)}]{bekenstein84}
\bibinfo{author}{\bibfnamefont{J.~D.} \bibnamefont{Bekenstein}}, in
  \emph{\bibinfo{booktitle}{Quantum Theory of Gravity}}, edited by
  \bibinfo{editor}{\bibfnamefont{S.~M.} \bibnamefont{Christensen}}
  (\bibinfo{publisher}{Adam Hilger}, \bibinfo{address}{Bristol},
  \bibinfo{year}{1984}).

\bibitem[{\citenamefont{Burgess}(2004)}]{burgess04}
\bibinfo{author}{\bibfnamefont{C.~P.} \bibnamefont{Burgess}},
  \bibinfo{journal}{Living Rev. Rel.} \textbf{\bibinfo{volume}{7}},
  \bibinfo{pages}{5} (\bibinfo{year}{2004}).

\bibitem[{\citenamefont{Donoghue}(1994)}]{donoghue94}
\bibinfo{author}{\bibfnamefont{J.~F.} \bibnamefont{Donoghue}},
  \bibinfo{journal}{Phys. Rev. D} \textbf{\bibinfo{volume}{50}},
  \bibinfo{pages}{3874} (\bibinfo{year}{1994}).

\bibitem[{\citenamefont{Donoghue}(1999)}]{donoghue97}
\bibinfo{author}{\bibfnamefont{J.~F.} \bibnamefont{Donoghue}}, in
  \emph{\bibinfo{booktitle}{The Eighth Marcel Grossmann Meeting}}, edited by
  \bibinfo{editor}{\bibfnamefont{T.}~\bibnamefont{Piran}} \bibnamefont{and}
  \bibinfo{editor}{\bibfnamefont{R.}~\bibnamefont{Ruffini}}
  (\bibinfo{publisher}{World Scientific}, \bibinfo{address}{Singapore},
  \bibinfo{year}{1999}), \eprint{gr-qc/9712070}.

\bibitem[{\citenamefont{Weinberg}()}]{weinberg05}
\bibinfo{author}{\bibfnamefont{S.}~\bibnamefont{Weinberg}},
  \eprint{hep-th/0506236}.

\bibitem[{\citenamefont{Hu and Verdaguer}(2003)}]{hu03a}
\bibinfo{author}{\bibfnamefont{B.~L.} \bibnamefont{Hu}} \bibnamefont{and}
  \bibinfo{author}{\bibfnamefont{E.}~\bibnamefont{Verdaguer}},
  \bibinfo{journal}{Class. Quant. Grav.} \textbf{\bibinfo{volume}{20}},
  \bibinfo{pages}{R1} (\bibinfo{year}{2003}).

\bibitem[{\citenamefont{Hu and Verdaguer}(2004)}]{hu04a}
\bibinfo{author}{\bibfnamefont{B.~L.} \bibnamefont{Hu}} \bibnamefont{and}
  \bibinfo{author}{\bibfnamefont{E.}~\bibnamefont{Verdaguer}},
  \bibinfo{journal}{Living Rev. Rel.} \textbf{\bibinfo{volume}{7}},
  \bibinfo{pages}{3} (\bibinfo{year}{2004}).

\bibitem[{\citenamefont{Roura and Verdaguer}()}]{roura03b}
\bibinfo{author}{\bibfnamefont{A.}~\bibnamefont{Roura}} \bibnamefont{and}
  \bibinfo{author}{\bibfnamefont{E.}~\bibnamefont{Verdaguer}},
  \bibinfo{note}{in preparation}.

\bibitem[{\citenamefont{Hu et~al.}(2004)\citenamefont{Hu, Roura, and
  Verdaguer}}]{hu04b}
\bibinfo{author}{\bibfnamefont{B.~L.} \bibnamefont{Hu}},
  \bibinfo{author}{\bibfnamefont{A.}~\bibnamefont{Roura}}, \bibnamefont{and}
  \bibinfo{author}{\bibfnamefont{E.}~\bibnamefont{Verdaguer}},
  \bibinfo{journal}{Phys. Rev. D} \textbf{\bibinfo{volume}{70}},
  \bibinfo{pages}{044002} (\bibinfo{year}{2004}).

\bibitem[{\citenamefont{Parentani}(2001{\natexlab{a}})}]{parentani01b}
\bibinfo{author}{\bibfnamefont{R.}~\bibnamefont{Parentani}},
  \bibinfo{journal}{Int. J. Theor. Phys.} \textbf{\bibinfo{volume}{40}},
  \bibinfo{pages}{2201} (\bibinfo{year}{2001}{\natexlab{a}}).

\bibitem[{\citenamefont{Parentani}(2002)}]{parentani02}
\bibinfo{author}{\bibfnamefont{R.}~\bibnamefont{Parentani}},
  \bibinfo{journal}{Int. J. Theor. Phys.} \textbf{\bibinfo{volume}{41}},
  \bibinfo{pages}{2175} (\bibinfo{year}{2002}).

\bibitem[{\citenamefont{Hu and Roura}()}]{hu06b}
\bibinfo{author}{\bibfnamefont{B.~L.} \bibnamefont{Hu}} \bibnamefont{and}
  \bibinfo{author}{\bibfnamefont{A.}~\bibnamefont{Roura}}, \bibinfo{note}{in
  preparation}.

\bibitem[{\citenamefont{Misner et~al.}(1973)\citenamefont{Misner, Thorne, and
  Wheeler}}]{misner73}
\bibinfo{author}{\bibfnamefont{C.~W.} \bibnamefont{Misner}},
  \bibinfo{author}{\bibfnamefont{K.~S.} \bibnamefont{Thorne}},
  \bibnamefont{and} \bibinfo{author}{\bibfnamefont{J.~A.}
  \bibnamefont{Wheeler}}, \emph{\bibinfo{title}{Gravitation}}
  (\bibinfo{publisher}{Freeman}, \bibinfo{address}{San Francisco},
  \bibinfo{year}{1973}).

\bibitem[{\citenamefont{Wald}(1984)}]{wald84}
\bibinfo{author}{\bibfnamefont{R.~M.} \bibnamefont{Wald}},
  \emph{\bibinfo{title}{General Relativity}} (\bibinfo{publisher}{The
  University of Chicago Press}, \bibinfo{address}{Chicago},
  \bibinfo{year}{1984}).

\bibitem[{\citenamefont{Page}(1976)}]{page76}
\bibinfo{author}{\bibfnamefont{D.~N.} \bibnamefont{Page}},
  \bibinfo{journal}{Phys. Rev. D} \textbf{\bibinfo{volume}{13}},
  \bibinfo{pages}{198} (\bibinfo{year}{1976}).

\bibitem[{\citenamefont{York}(1985)}]{york85}
\bibinfo{author}{\bibfnamefont{J.~W.} \bibnamefont{York}},
  \bibinfo{journal}{Phys. Rev. D} \textbf{\bibinfo{volume}{31}},
  \bibinfo{pages}{775} (\bibinfo{year}{1985}).

\bibitem[{\citenamefont{Candelas}(1980)}]{candelas80}
\bibinfo{author}{\bibfnamefont{P.}~\bibnamefont{Candelas}},
  \bibinfo{journal}{Phys. Rev. D} \textbf{\bibinfo{volume}{21}},
  \bibinfo{pages}{2185} (\bibinfo{year}{1980}).

\bibitem[{\citenamefont{Page}(1982)}]{page82}
\bibinfo{author}{\bibfnamefont{D.~N.} \bibnamefont{Page}},
  \bibinfo{journal}{Phys. Rev. D} \textbf{\bibinfo{volume}{25}},
  \bibinfo{pages}{1499} (\bibinfo{year}{1982}).

\bibitem[{\citenamefont{Howard and Candelas}(1984)}]{howard84a}
\bibinfo{author}{\bibfnamefont{K.~W.} \bibnamefont{Howard}} \bibnamefont{and}
  \bibinfo{author}{\bibfnamefont{P.}~\bibnamefont{Candelas}},
  \bibinfo{journal}{Phys. Rev. Lett.} \textbf{\bibinfo{volume}{53}},
  \bibinfo{pages}{403} (\bibinfo{year}{1984}).

\bibitem[{\citenamefont{Howard}(1984)}]{howard84b}
\bibinfo{author}{\bibfnamefont{K.~W.} \bibnamefont{Howard}},
  \bibinfo{journal}{Phys. Rev. D} \textbf{\bibinfo{volume}{30}},
  \bibinfo{pages}{2532} (\bibinfo{year}{1984}).

\bibitem[{\citenamefont{Anderson et~al.}(1995)\citenamefont{Anderson, Hiscock,
  and Samuel}}]{anderson95}
\bibinfo{author}{\bibfnamefont{P.~R.} \bibnamefont{Anderson}},
  \bibinfo{author}{\bibfnamefont{W.~A.} \bibnamefont{Hiscock}},
  \bibnamefont{and} \bibinfo{author}{\bibfnamefont{D.~A.}
  \bibnamefont{Samuel}}, \bibinfo{journal}{Phys. Rev. D}
  \textbf{\bibinfo{volume}{51}}, \bibinfo{pages}{4337} (\bibinfo{year}{1995}).

\bibitem[{\citenamefont{Trivedi}(1993)}]{trivedi93}
\bibinfo{author}{\bibfnamefont{S.~P.} \bibnamefont{Trivedi}},
  \bibinfo{journal}{Phys. Rev. D} \textbf{\bibinfo{volume}{47}},
  \bibinfo{pages}{4233} (\bibinfo{year}{1993}).

\bibitem[{\citenamefont{Strominger and Trivedi}(1993)}]{strominger93}
\bibinfo{author}{\bibfnamefont{A.}~\bibnamefont{Strominger}} \bibnamefont{and}
  \bibinfo{author}{\bibfnamefont{S.~P.} \bibnamefont{Trivedi}},
  \bibinfo{journal}{Phys. Rev. D} \textbf{\bibinfo{volume}{48}},
  \bibinfo{pages}{5778} (\bibinfo{year}{1993}).

\bibitem[{\citenamefont{Lombardo et~al.}(1999)\citenamefont{Lombardo,
  Mazzitelli, and Russo}}]{lombardo99}
\bibinfo{author}{\bibfnamefont{F.~C.} \bibnamefont{Lombardo}},
  \bibinfo{author}{\bibfnamefont{F.~D.} \bibnamefont{Mazzitelli}},
  \bibnamefont{and} \bibinfo{author}{\bibfnamefont{J.~G.} \bibnamefont{Russo}},
  \bibinfo{journal}{Phys. Rev. D} \textbf{\bibinfo{volume}{59}},
  \bibinfo{pages}{064007} (\bibinfo{year}{1999}).

\bibitem[{\citenamefont{Parentani}(2001{\natexlab{b}})}]{parentani01a}
\bibinfo{author}{\bibfnamefont{R.}~\bibnamefont{Parentani}},
  \bibinfo{journal}{Phys. Rev. D} \textbf{\bibinfo{volume}{63}},
  \bibinfo{pages}{041503} (\bibinfo{year}{2001}{\natexlab{b}}).

\bibitem[{\citenamefont{Campos and Hu}(1998)}]{campos98}
\bibinfo{author}{\bibfnamefont{A.}~\bibnamefont{Campos}} \bibnamefont{and}
  \bibinfo{author}{\bibfnamefont{B.~L.} \bibnamefont{Hu}},
  \bibinfo{journal}{Phys. Rev. D} \textbf{\bibinfo{volume}{58}},
  \bibinfo{pages}{125021} (\bibinfo{year}{1998}).

\bibitem[{\citenamefont{Campos and Hu}(1999)}]{campos99}
\bibinfo{author}{\bibfnamefont{A.}~\bibnamefont{Campos}} \bibnamefont{and}
  \bibinfo{author}{\bibfnamefont{B.~L.} \bibnamefont{Hu}},
  \bibinfo{journal}{Int. J. Theor. Phys.} \textbf{\bibinfo{volume}{38}},
  \bibinfo{pages}{1253} (\bibinfo{year}{1999}).

\bibitem[{\citenamefont{Ford and Roman}(2005)}]{ford05}
\bibinfo{author}{\bibfnamefont{L.~H.} \bibnamefont{Ford}} \bibnamefont{and}
  \bibinfo{author}{\bibfnamefont{T.~A.} \bibnamefont{Roman}},
  \bibinfo{journal}{Phys. Rev. D} \textbf{\bibinfo{volume}{72}},
  \bibinfo{pages}{105010} (\bibinfo{year}{2005}).

\bibitem[{\citenamefont{Phillips and Hu}(2001)}]{phillips01}
\bibinfo{author}{\bibfnamefont{N.~G.} \bibnamefont{Phillips}} \bibnamefont{and}
  \bibinfo{author}{\bibfnamefont{B.~L.} \bibnamefont{Hu}},
  \bibinfo{journal}{Phys. Rev. D} \textbf{\bibinfo{volume}{63}},
  \bibinfo{pages}{104001} (\bibinfo{year}{2001}).

\bibitem[{\citenamefont{Phillips and Hu}(2003)}]{phillips03}
\bibinfo{author}{\bibfnamefont{N.~G.} \bibnamefont{Phillips}} \bibnamefont{and}
  \bibinfo{author}{\bibfnamefont{B.~L.} \bibnamefont{Hu}},
  \bibinfo{journal}{Phys. Rev. D} \textbf{\bibinfo{volume}{67}},
  \bibinfo{pages}{104002} (\bibinfo{year}{2003}).

\bibitem[{\citenamefont{Gibbons and Hawking}(1977)}]{gibbons77a}
\bibinfo{author}{\bibfnamefont{G.~W.} \bibnamefont{Gibbons}} \bibnamefont{and}
  \bibinfo{author}{\bibfnamefont{S.~W.} \bibnamefont{Hawking}},
  \bibinfo{journal}{Phys. Rev. D} \textbf{\bibinfo{volume}{15}},
  \bibinfo{pages}{2738} (\bibinfo{year}{1977}).

\bibitem[{\citenamefont{Barrab\`{e}s et~al.}(1999)\citenamefont{Barrab\`{e}s,
  Frolov, and Parentani}}]{barrabes99}
\bibinfo{author}{\bibfnamefont{C.}~\bibnamefont{Barrab\`{e}s}},
  \bibinfo{author}{\bibfnamefont{V.}~\bibnamefont{Frolov}}, \bibnamefont{and}
  \bibinfo{author}{\bibfnamefont{R.}~\bibnamefont{Parentani}},
  \bibinfo{journal}{Phys. Rev. D} \textbf{\bibinfo{volume}{59}},
  \bibinfo{pages}{124010} (\bibinfo{year}{1999}).

\bibitem[{\citenamefont{Barrab\`{e}s et~al.}(2000)\citenamefont{Barrab\`{e}s,
  Frolov, and Parentani}}]{barrabes00}
\bibinfo{author}{\bibfnamefont{C.}~\bibnamefont{Barrab\`{e}s}},
  \bibinfo{author}{\bibfnamefont{V.}~\bibnamefont{Frolov}}, \bibnamefont{and}
  \bibinfo{author}{\bibfnamefont{R.}~\bibnamefont{Parentani}},
  \bibinfo{journal}{Phys. Rev. D} \textbf{\bibinfo{volume}{62}},
  \bibinfo{pages}{044020} (\bibinfo{year}{2000}).

\end{thebibliography}

\end{document}